\documentclass[]{jpsj2}

\def\dsum#1#2{\displaystyle\sum_{#1}^{#2}}
\def\dfrac#1#2{\displaystyle\frac{#1}{#2}}
\def\<{\langle}
\def\>{\rangle}
\def\({\left(}
\def\){\right)}
\def\[{\left[}
\def\]{\right]}
\def\mibs#1{\mbox{\boldmath $\scriptstyle{#1}$}}
\def\up{\uparrow}
\def\dn{\downarrow}
\def\e{\mathrm{e}}
\def\s{\sigma}
\def\i{\mathrm{i}}

\def\lco{\mathrm{La}_2\mathrm{CuO}_4}
\def\lsco{\mathrm{La}_{2-x}\mathrm{Sr}_{x}\mathrm{CuO}_4}

\def\runtitle{
Fermi Arc of Metallic Diagonal Stripes in High $T_\mathrm{c}$ Cuprates
}
\def\runauthor{
Eiji {\sc Kaneshita}, Masanori {\sc Ichioka} and Kazushige {\sc Machida}
}
\title{
Fermi Arc of Metallic Diagonal Stripes in High $T_\mathrm{c}$ Cuprates
}
\author{
Eiji {\sc Kaneshita}\footnote{Present address: MS-B262, T-11, Los Alamos National Laboratory, Los Alamos, NM 87545, USA; E-mail: eiji@viking.lanl.gov},
Masanori {\sc Ichioka} and Kazushige {\sc Machida}
}
\inst
{Department of Physics, Okayama University, Okayama 700-8530}
\recdate{\today}
\abst{
Spectral weight is investigated for metallic diagonal stripe state
in two dimensional Hubbard model,
and Fermi arc observed by angle-resolved photoemission spectroscopy on $\lsco$ is discussed.
The Fermi arc coming from the mid-gap state of diagonal stripe
appears near $(\frac{\pi}{2},\frac{\pi}{2})$ and equivalent position in
the reciprocal space, and the gap opens below the mid-gap state.
We show how these spectral weight structure depends on the phasing
of stripes, i.e., site-centered or bond-centered stripes.
}
\kword{
Incommensurate spin-density wave,
Spectral weight,
Hubbard model,
High Tc superconductor
}
\begin{document}
\sloppy
\maketitle


Much attention has been focused on high $T_\mathrm{c}$
cuprate superconductors since its discovery over 15 years ago.
The superconducting pairing mechanism is not cleared yet.
Superconductivity in $\lsco$ (LSCO) is derived by hole doping
from the undoped parent compound $\lco$, which
is an antiferromagnetic (AF) quantum magnet.
It is expected that an understanding of behaviors of
doped holes in two dimensional CuO$_2$ plane is a key to high
$T_\mathrm{c}$ mechanism.
In order to describe doped holes, the stripe picture has been
proposed\cite{machida1,kato,zaanen,poiblanc,schulz} and
developed\cite{tranquada,emery,salkola,white,machida2,ichioka1,ichioka2,kaneshita1,kaneshita2},
where one-dimensional spin and charge arrangement is characterized
by incommensurate wave vector $\mib{Q}$.
A variety of experiments support the stripe picture now\cite{kivelson}.
Since there are many kinds of high $T_\mathrm{c}$ cuprates
which show the stripe orderings, investigating the electronic states
in stripes is beneficial to understanding
the mechanism of the high $T_\mathrm{c}$ superconductivity.

The static incommensurate spin order is observed by
elastic neutron scattering experiment,
which suggests that the vertical stripe structure with ordering vectors $\mib{Q}=2\pi(\frac{1}{2}\pm\delta,\frac{1}{2})$ or
$\mib{Q}=2\pi(\frac{1}{2},\frac{1}{2}\pm\delta)$ in the superconductor
phase ($x>0.05$) changes to the diagonal stripe
with ordering vectors
$\mib{Q}=2\pi(\frac{1}{2}\pm\delta,\frac{1}{2}\pm\delta)$ or
$\mib{Q}=2\pi(\frac{1}{2}\pm\delta,\frac{1}{2}\mp\delta)$
in the insulator phase ($x<0.05$).\cite{Matsuda}
It is noted that the ratio $\delta/x$ changes
at the superconductor-insulator transition.
In the superconductor phase, $\delta \sim x$, which means that
the hole density is 0.5 per stripe, suggesting the metallic behavior.\cite{tranquada}
On the other hand, In the insulator phase, $\delta \sim \frac{x}{2}$, which means that the hole density is 1 per stripe, suggesting the insulating behavior with full-filled holes.
In the theoretical analysis of the electronic state,
there appears the mid-gap state for the stripe states.\cite{machida2,ichioka1}
In the case $\delta = \frac{x}{2}$,
the Fermi level is located in the gap between the mid-gap state
and the lower band, suggesting the insulating state.
When $\delta > \frac{x}{2}$, it becomes metallic state, because the  Fermi level is located in the mid-gap state.
The transition from the vertical metallic stripe state to the insulating diagonal stripe was understood within this analysis.\cite{machida2,ichioka1}

Insulator phase of underdoped LSCO for $0.02<x<0.05$, existing
between AF phase  and superconductor phase, still
exhibits a full of mysteries up to now.
Previously, this phase is considered as spin-glass, and now
as diagonal stripe phase.
Recently, the insulating character of this phase is re-examined
by the recent experiments of
angle-resolved photoemission spectroscopy (ARPES)~\cite{Yoshida}
and transport studies~\cite{Ando} on LSCO,
which suggest that the diagonal stripe phase is ``metallic''.

The ARPES experiments show some interesting and important results
on the electronic state of low-doped LSCO, such as
the nodal Fermi surface, the Fermi-surface arc;
the kink structure,
and so on.~\cite{Yoshida,shen}
For lower doping ($x\sim0.03$), especially,
the spectral weight at Fermi level appears only around
$(\frac{\pi}{2},\frac{\pi}{2})$  in the reciprocal space.~\cite{Yoshida}
This Fermi level state, called as ``Fermi arc'', means metallic character of this state, while it has been considered as an insulator phase.
Furthermore, below the Fermi level, the electronic dispersion has a gap which separates the state at the Fermi level from the deeper dispersion curve.
This is a unique character of LSCO, and
suggests that the mid-gap state touches across with the Fermi level.
The metallic behavior is also supported by the recent transport studies
on LSCO, which indicate metallic behaviors at high temperatures,
even down to the extremely light doping ($x \sim 0.01$).\cite{Ando}
There, the insulator behavior at low temperature is considered as
the localization effect.

On the other hand, when we see the neutron scattering data~\cite{Matsuda} for low-doped LSCO based on the above-mentioned picture more closely,
it is important to notice that the incommensurability $\delta$ slightly deviates from the expected insulating line $\delta=\frac{x}{2}$,
suggesting the metallic behavior due to the mid-gap state touching the
Fermi level.

In this paper, we study the spectral weight of the metallic
diagonal stripe state assuming the deviation from the relation
$\delta=\frac{x}{2}$, and discuss the Fermi arc state observed by
ARPES in the standpoint of the stripe picture.
We also analyze the relation of the spectral weight and
the phasing of the stripe, i.e.,
bond-centered or site-centered stripes.
We use simple Hubbard model for this study
to extract essential features of the stripe state
without material-dependent details.


We start with the Hubbard model on a two dimensional square lattice,
\begin{eqnarray}
H=-\sum_{i,j,\s}t_{i,j}C^{\dagger}_{i,\s}
C_{j,\s}+U\sum_in_{i\uparrow}n_{i\downarrow},
\label{eq:Hamiltonian1}
\end{eqnarray}
where $\sigma$ is a spin index and $i=(i_x,i_y)$ is a site index.
$t_{i,j}=t$ ($t'$) for the (next) nearest neighbor sites $i$ and $j$.
In this article, we set the parameters: $U=5.0t, t'=-0.2t$.
We assume a periodic spin and associated charge orderings
to take into account the stripes, introducing the order parameters
\begin{eqnarray}
\< n_{i,\s} \>
=\dsum{l=0,1,2,\cdots,N-1}{} \e^{\i l\mibs{Q}\cdot{\mibs{r}_i}}
\< n_{l\mibs{Q},\s} \>
\label{eq:average}
\end{eqnarray}
with $\delta=1/N$.
The associated spin density $S_{z,i}$ and charge density $n_i$ are,
respectively, given by
\begin{eqnarray}
S_{z,i}= \frac{1}{2}\(\, \< n_{i,\up} \>- \< n_{i,\dn} \> \,\),
\quad
n_{i}= \< n_{i,\up} \>+ \< n_{i,\dn} \> .
\label{eq:magnetization}
\end{eqnarray}
In $N$-site periodic case, the Brillouin zone is reduced to $1/N$-area.
The energy dispersion is split to $N$ bands.
We write $\mib{k}=\mib{k}_0 + m \mib{Q}$ ($m=0,\ 1,\ \cdots,\ N-1$),
where $\mib{k}_0$ is restricted within the  reduced Brillouin zone.
Then the Hamiltonian is reduced to
\begin{eqnarray}
&&H=\sum_{\mibs{k}_0,\s}
\sum_{m,n}C^{\dagger}_{\mibs{k}_0+m\mibs{Q},\s}
({\hat H}_{\mibs{k}_0,\s})_{m,m'}C_{\mibs{k}_0+m'\mibs{Q},\s}
\nonumber \\
&&=\sum_{\mibs{k}_0,\s,\alpha}E_{\mibs{k}_0,\s,\alpha}
\gamma^\dagger_{\mibs{k}_0,\s,\alpha}
\gamma_{\mibs{k}_0,\s,\alpha}.
\label{eq:HamiltonianMF}
\end{eqnarray}
The $N \times N$ Hamiltonian matrix
\begin{eqnarray}
\( {\hat H}_{\mibs{k}_0,\s}  \)_{m,m'}
=\epsilon(\mib{k}_0+m\mib{Q})\delta_{m,m'}
+U \langle n_{(m-m')\mibs{Q},-\s} \rangle
\nonumber \\
\label{eq:matrix}
\end{eqnarray}
is diagonalized by a unitary transformation:
\begin{eqnarray}
C_{\mibs{k}_0+m\mibs{Q},\s}=\sum_{\alpha}
u_{\mibs{k}_0,\s,\alpha,m}\gamma_{\mibs{k}_0,\s,\alpha}.
\label{eq:unitary}
\end{eqnarray}
In eq. (\ref{eq:matrix}),
$\epsilon(\mib{k})=-2t(\cos k_x +\cos k_y) -4t'\cos k_x \cos k_y$.
The calculation is iterated until all the order parameters satisfy
the self-consistent condition:
\begin{eqnarray}
\langle n_{l\mibs{Q}\s} \rangle = N_k^{-1}
\sum_{\mibs{k}_0, m, \alpha} u^{\ast}_{\mibs{k}_0,\s,\alpha, m}
u_{\mibs{k}_0,\s,\alpha, m+l}f(E_{\mibs{k}_0,\s, \alpha})
\nonumber \\
\label{eq:averageN}
\end{eqnarray}
with $N_k=\sum_{\mibs{k}_0,m}1$.

Using the selfconsistently obtained wave functions,
we construct the thermal Green's function as
\begin{eqnarray}
g_\s(\mib{r},\mib{r}', \i \omega_n)=\sum_{\mibs{k}_0,\alpha}
\frac{u_{\mibs{k}_0,\s,\alpha}(\mib{r})
u^\ast_{\mibs{k}_0,\s,\alpha}(\mib{r'})}
{ \i \omega_n - E_{\mibs{k}_0,\s,\alpha}}
\label{eq:Gf}
\end{eqnarray}
with
\begin{eqnarray}
u_{\mibs{k}_0,\s,\alpha}(\mib{r}_i)=N_k^{-1/2} \sum_m
{\rm e}^{\i(\mibs{k}_0+m\mibs{Q})\cdot \mibs{r}_i}
u_{\mibs{k}_0,\s,\alpha,m}.
\label{eq:wavefn}
\end{eqnarray}
The spectral weight is given by
$N(\mib{k},E)=N_\up(\mib{k},E)+N_\dn(\mib{k},E)$ with
\begin{eqnarray}
&&\hspace{-1.0cm}N_\s(\mib{k},E)
=-\dfrac{1}{\pi}\,\mathrm{Im}\,G_{\s}(\mib{k},E)\nonumber\\
&&\hspace{-1.0cm}=\dsum{\mibs{k}_0,\alpha,m}{}|u_{\mibs{k}_0,\s,\alpha,m}|^2
\,\delta\(E-E_{\mibs{k}_0,\alpha,m}\)
\,\delta\(\mib{k}_0+m\mib{Q}-\mib{k}\),
\label{eq:spctrl}
\end{eqnarray}

where $G_\s(\mib{k},E)$ is a Fourier transformation of
the retarded Green's function:
\begin{eqnarray}
&&\hspace{-1cm}G_\s(\mib{k},E)
=N_k^{-1}\dsum{\mibs{r},\mibs{r}'}{}\e^{-\i\mib{k}\(\mib{r}-\mib{r}'\)}
g_\s(\mib{r},\mib{r}', E-\i 0^+). \qquad
\label{eq:rGF}
\end{eqnarray}

We study the metallic diagonal stripe with the ordering vector
$\mib{Q}=2\pi(\frac{1}{2}\pm\delta, \frac{1}{2}\pm\delta)$.
When spin structure has $N$-site periodicity, $\delta=1/N$.
We show results of numerical calculations for two cases, i.e.,
for 64-site periodic case with $x=0.03$, and
for 40-site periodic case with $x =0.049 \sim 0.047$.
These are slightly deviated from the insulator relation $\delta = \frac{x}{2}$.

\begin{figure}
\begin{center}
\includegraphics[width=0.8\linewidth, bb=153 455 463 684]{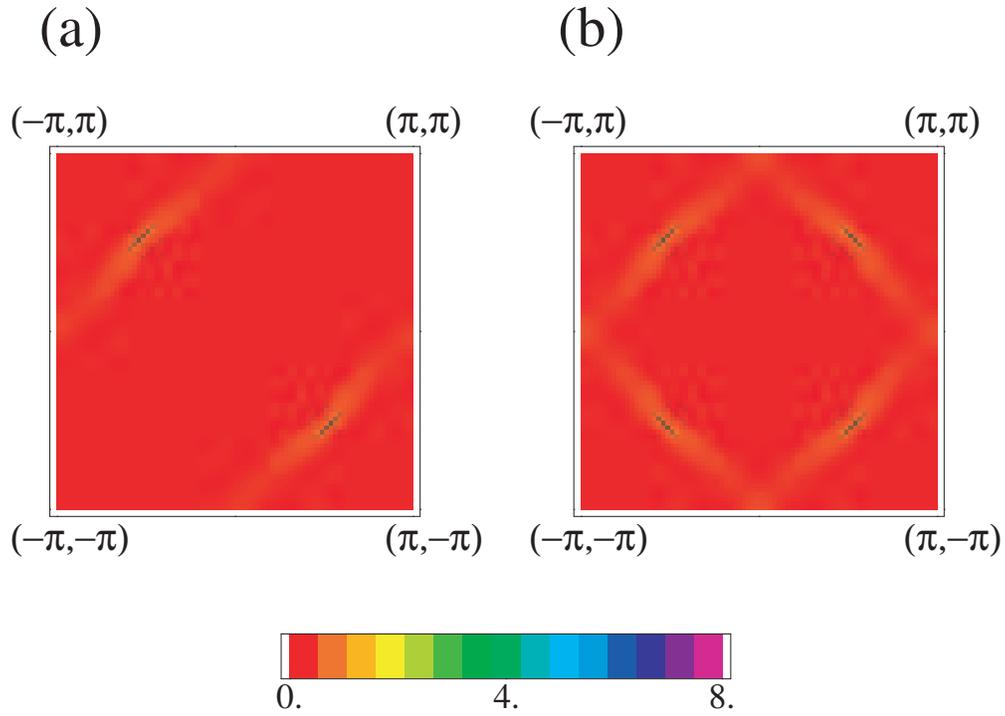}
\caption{(a) Density plot of the spectral weight at Fermi level $N(\mib{k}, E_{F}) $ for $x=0.03$ and $2 \delta =0.03125$ ($N=64$).
This shows the Fermi surface state of the metallic diagonal stripe.
The region of $2\pi \times 2\pi$ reciprocal space is presented.
(b) The combination of $N(\mib{k},E_\mathrm{F})$ from two domains is drawn, i.e.,
the domain with
$\mib{Q}=2\pi(\frac{1}{2}\pm\delta, \frac{1}{2}\pm\delta)$ and
the $90^\circ$-rotated one with
$\mib{Q}=2\pi(\frac{1}{2}\mp\delta, \frac{1}{2}\pm\delta)$.
}
\label{fig:fermi1}
\end{center}
\end{figure}

The spectral weight at Fermi level $N(\mib{k},E_\mathrm{F})$
is shown in Fig. \ref{fig:fermi1}(a) for $x=0.03$.
There, the one-dimensional Fermi surface appears
around $(-\pi/2,\pi/2)$ and $(\pi/2,-\pi/2)$.
Roughly speaking, the Fermi surface in the diagonal stripe tends to become diagonal one-dimensional one,
reflecting the one-dimensional conduction
along the diagonal stripe line.
We note that the spectral weight on the one-dimensional
Fermi surface is large only near  $(-\pi/2,\pi/2)$ and $(\pi/2,-\pi/2)$,
where original Fermi surface is located in the case of no stripe order.
Therefore, the Fermi level state is observed as the
Fermi arc near $(-\pi/2,\pi/2)$ and $(\pi/2,-\pi/2)$.

When our result of the Fermi level state is compared with
the ARPES data\cite{Yoshida}, we should take into account the domain
structure of the stripe state.
It is considered that two types of domains, i.e.,
the domain with
$\mib{Q}=2\pi(\frac{1}{2}\pm\delta, \frac{1}{2}\pm\delta)$ and
the $90^\circ$-rotated one with
$\mib{Q}=2\pi(\frac{1}{2}\mp\delta, \frac{1}{2}\pm\delta)$
co-exist in LSCO.
When the contributions from these domains are combined,
the spectral weight is given by
$N(\mib{k},E_\mathrm{F})+N(k_x \leftrightarrow k_y,E_\mathrm{F})$,
as shown in Fig. \ref{fig:fermi1}(b).
This seems rather similar to the results of ARPES, where
four Fermi arcs appears near $(\pi/2,\pi/2)$ and equivalent positions
in $2\pi \times 2\pi$ reciprocal space.

\begin{figure}
\begin{center}
\includegraphics[width=0.8\linewidth]{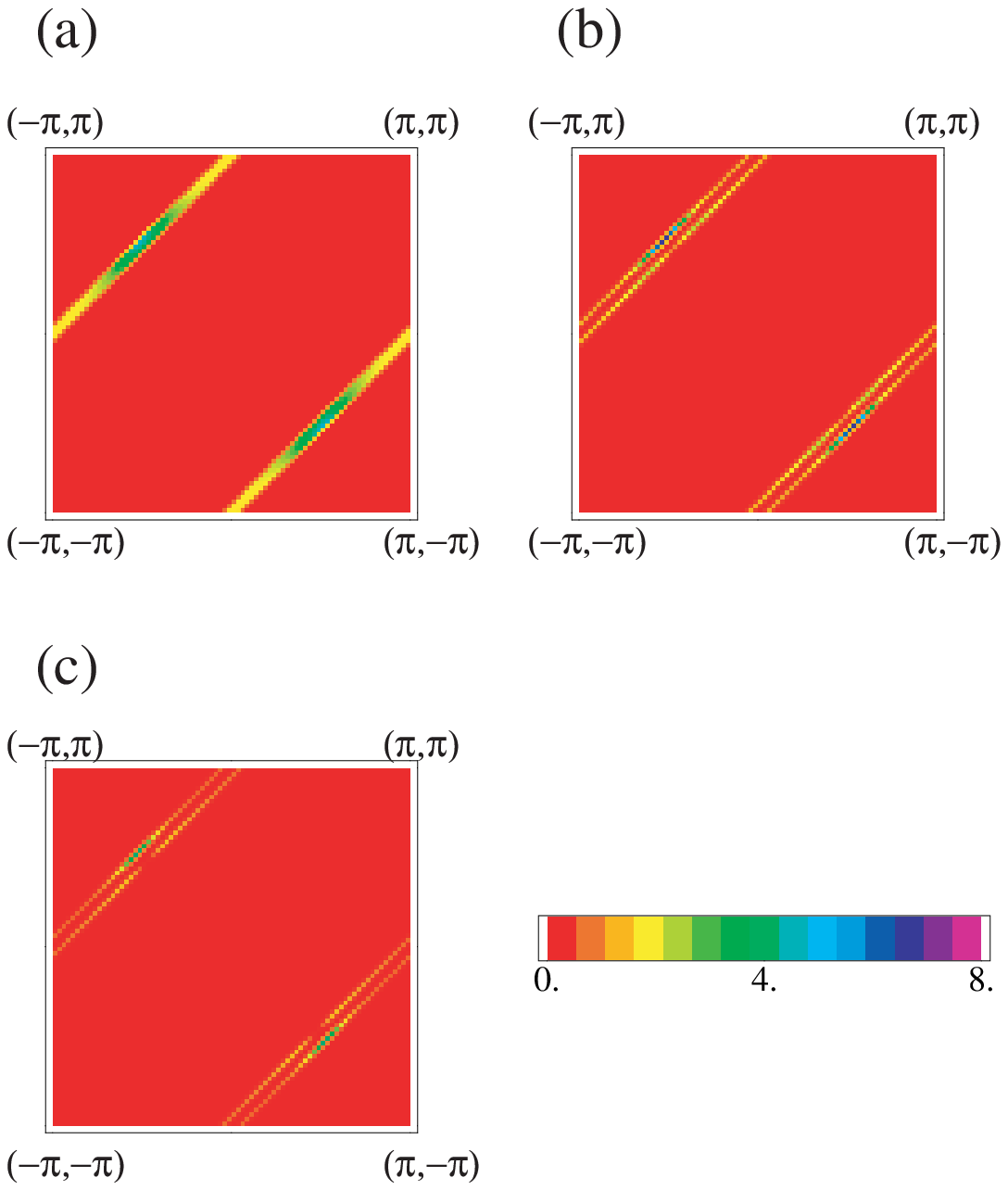}
\caption{
Density plot of the spectral weight at Fermi level
$N(\mib{k},E_\mathrm{F})$ for $2\delta=0.05$ ($N=40$).
(a) $x=0.049$, (b) $x=0.048$ and (c) $x=0.047$.
}
\label{fig:fermi2}
\end{center}
\end{figure}

The spectral weight for the 40 site periodicity case ($2\delta=0.05$)
is shown in Fig. \ref{fig:fermi2}.
In this case, we consider the deviation from the insulating case, namely, $(2\delta -x)$-dependence by changing the filling $x$.
For $x=0.049$ (Fig. \ref{fig:fermi2}(a)), the Fermi arc state has large spectral weight near $(-\pi/2,\pi/2)$ and $(\pi/2,-\pi/2)$,
and also has straight tails toward $(\pm\pi,0)$ and $(0,\pm\pi)$.
This Fermi arc tail structure is different from that in Fig. \ref{fig:fermi1}, while $2\delta-x \sim 0.01$ in both cases.
By comparing Figs. \ref{fig:fermi2}(a)-(c),
we see that the intensity of tails around $(\pm\pi,0)$ and $(0,\pm\pi)$ vanishes gradually with reducing hole-concentration $x$, i.e. the difference between $2\delta$ and $x$ increases.
The spectral weight for $x=0.047$ (Fig. \ref{fig:fermi2} (c)) is similar to the case for $x=0.03$ (Fig. \ref{fig:fermi1}).

\begin{figure}
\begin{center}
\includegraphics[width=0.8\linewidth]{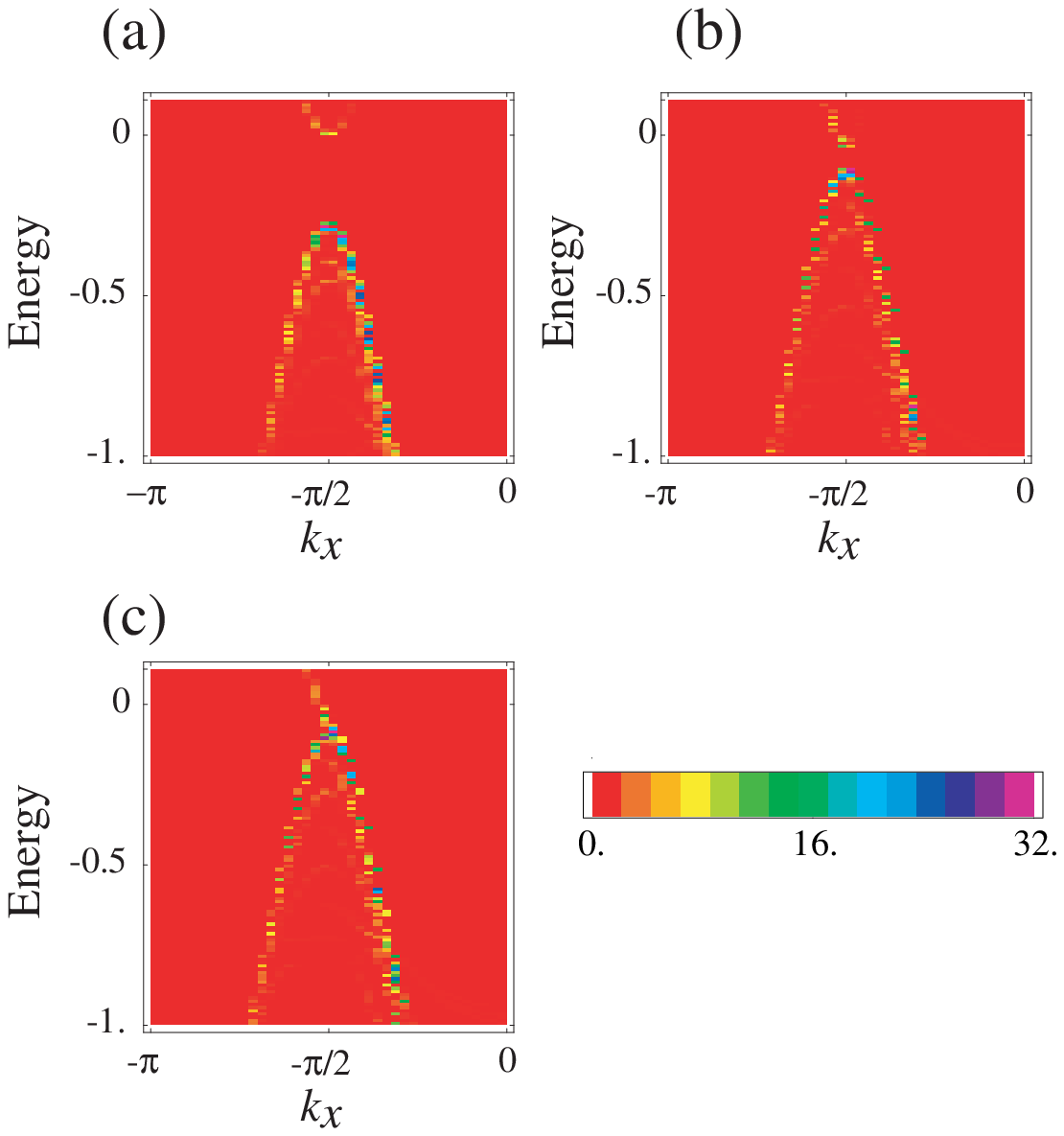}
\caption{
Density plot of the spectral weight $N(\mib{k},E)$ for $2\delta=0.05$
along the diagonal path $\mib{k}$: $(-\pi,\pi)$ -- $(0,0)$.
(a) $x=0.049$, (b) $x=0.048$, (c) $x=0.047$.
}
\label{fig:spctrl}
\end{center}
\end{figure}

Figures. \ref{fig:spctrl}(a)--(c) show the spectral weight
along the path $\mib{k}$: $(-\pi,\pi)$ -- $(0,0)$ for hole-concentration
$x=0.049$, 0.048, 0.047, respectively.
In the stripe state, the dispersion of the original band
$\epsilon(\mib{k})$ is folded and split to $N$ bands
by $1/N$-reduction of the Brillouin zone,
and some gaps open.
It is noted that the spectral weight on each band reflects
the strength of the mixing with the original band $\epsilon(\mib{k})$.
In the commensurate AF state, the original band is divided to the upper and
lower bands by opening the AF gap.
In the incommensurate AF state for stripes, there appears the mid gap state within the main AF gap.\cite{machida2,ichioka1}
In Fig. \ref{fig:spctrl}(a), the Fermi level state at zero-energy is the mid-gap state,
which is isolated from the upper and lower bands.
Below the mid-gap state, we see the gap and the lower band.
This mid-gap state and the gap structure is resemble to
the spectral weight observed by ARPES on LSCO.

The gap structure drastically changes with changing $x$.
The gap is reduced as the difference $2\delta-x$ increase,
and vanishes for $x=0.047$.
For $x=0.047$, the mid-gap band overlaps with the upper and
lower bands, and the gap vanishes in the density of states.
This drastic change is not only due to the shift of the chemical potential, but also the change of the stripe structure as is seen below.

\begin{figure}
\begin{center}
\includegraphics[width=0.8\linewidth]{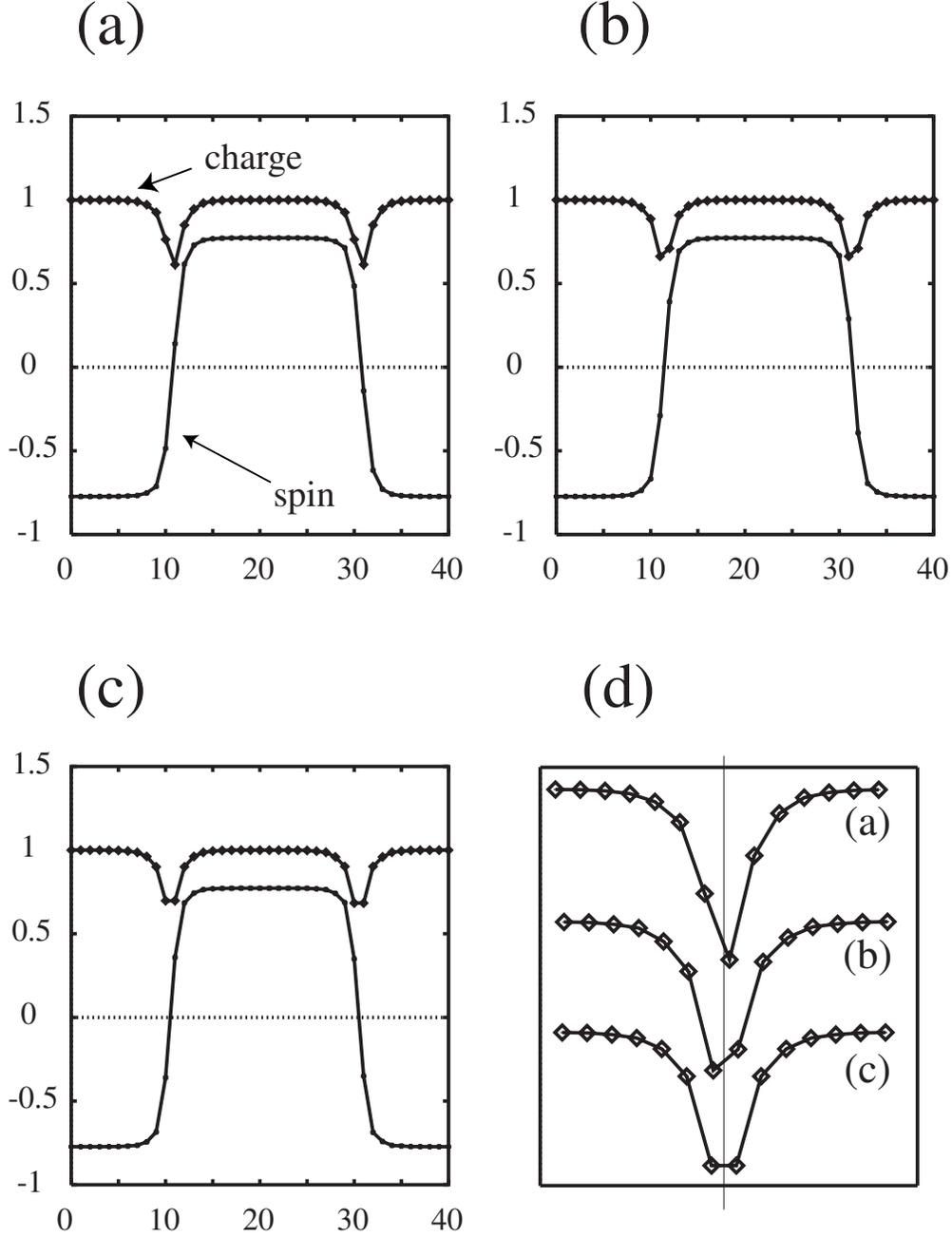}
\caption{
Spin density $(-1)^{i_x+i_y}S_{z,i}$ and charge density $n_i$ for
$x=0.049$ (a), $0.048$ (b) and  $0.047$ (c).
Horizontal axis shows the site-index.
The incommensurability is fixed at $2\delta=0.05$.
(d) Charge structures around the center of stripe in (a)-(c) are
magnified to see the $x$-dependence of the structure.
The vertical dotted line shows the center of the stripe i.e. the domain wall.
For $x=0.047$, the domain wall is located in a bond center.
For $x=0.049$ and $x=0.048$, it is away from the bond center.
}
\label{fig:density}
\end{center}
\end{figure}

To investigate differences among these three cases in more detail, let us examine the spin- and charge-density structure in real space, for each case.
As shown in Fig. \ref{fig:density}, the stripe structure varies from the site-centered one to the bond-centered one, as $x$ changes away from the insulator relation $2\delta=x$.
For an insulating diagonal case when $2\delta=x$,
the site-centered stripe is obtained in our calculation.
The relation of the spectral weight and the phasing of the stripe
is as follows.
In the site-centered stripe case, the gap below the mid-gap state
is large (Fig. \ref{fig:spctrl}(a)), and the Fermi arc has long straight tail toward $(\pm\pi,0)$ and $(0,\pm\pi)$ (Fig. \ref{fig:fermi2}(a)).
In the bond-centered stripe case, the gap width decreases (Figs. \ref{fig:spctrl}(b) and (c)), and the Fermi arc is localized only around
$(-\pi/2,\pi/2)$ and $(\pi/2,-\pi/2)$ (Figs. \ref{fig:fermi2}(b) and (c)).
The spectral weight in Fig. \ref{fig:fermi2}(a) has large intensity,
because the band bottom of the mid-gap state touches the Fermi energy
and the density of states becomes large at the band bottom
due to the van-Hove singularity.

For the previous $x=0.03$ case,
we obtained the almost bond-centered stripe structure,
and small gap below the mid-gap state.
That is, the results for the $x=0.03$ case is almost same
as in the $x=0.047 \sim 0.048$ case.
For $x=0.03$, the stripe becomes bond-centered quickly,
while the difference $2\delta-x$ is still small.

In summary,
we calculated the spectral weight and stripe structure for metallic diagonal stripes assuming the deviation from the insulator relation $\delta=\frac{x}{2}$.
In this calculation, we have shown that the Fermi arc appears near $(\pm\pi/2,\pm\pi/2)$ and $(\mp\pi/2,\pm\pi/2)$ as a result of the mid-gap state in the metallic stripe state, and the gap exists below the mid-gap state in the dispersion relation.
Therefore, the Fermi arc state observed ARPES on LSCO can be qualitatively reproduced by the stripe picture.
As for the phasing of the stripe, the stripe structure varies from site- to bond-centered one, as the deviation from the insulator relation $2\delta-x$ increases towards the metallic relation $\delta=x$.
These structure differences of the site- or bond-centered stripe affects the spectral weight, such as the gap width below the mid-gap state, or the extention of the tail of the Fermi arc.

 Since our calculation assumes the static stripe structure,
we obtain straight Fermi arcs coming from the one-dimensional
Fermi surface.
If we consider the effects of fluctuation or disorder
of the stripes, one-dimensional Fermi surface will be smeared,
and some spectral weight will appear on the Fermi surface of the original dispersion $\epsilon(\mib{k})$, as discussed in vertical stripe case.~\cite{salkola}
In this situation, we expect the Fermi arc will be curving shape
along the original Fermi surface.
Study on these effects remains to be a future problem.

We thank T. Yoshida and A. Fujimori for valuable information and discussion.



\end{document}